# The interplay between ferroelectricity and electrochemical reactivity on the surface of binary ferroelectric $Al_xB_{1-x}N$

Yongtao Liu,[1*] Anton Ievlev,[1] Joseph Casamento,[2] John Hayden,[2] Susan Trolier-McKinstry,[2] Jon-Paul Maria,[2] Sergei V. Kalinin,[3**] and Kyle P. Kelley[1***]

[1] Center for Nanophase Materials Sciences, Oak Ridge National Laboratory, Oak Ridge, TN 37830, USA

[2] Materials Research Institute, Pennsylvania State University, University Park, PA 16802, USA

[3] Department of Materials Science and Engineering, University of Tennessee, Knoxville, TN 37996, USA

Corresponding emails: *liuy3@ornl.gov; **sergei2@utk.edu; ***kelleykp@ornl.gov

**Abstract:**

Polarization dynamics and domain structure evolution in ferroelectric $Al_{0.93}B_{0.07}N$ are studied using piezoresponse force microscopy and spectroscopies in ambient and controlled atmosphere environments. The application of negative unipolar, and bipolar first-order reverse curve (FORC) waveforms leads to a protrusion-like feature on the $Al_{0.93}B_{0.07}N$ surface and reduction of electromechanical response due to electrochemical reactivity. A surface change is also observed on the application of fast alternating current bias. At the same time, the application of positive biases does not lead to surface changes. Comparatively in a controlled glove box atmosphere, stable polarization patterns can be observed, with minuscule changes in surface morphology. This surface morphology change is not isolated to applying biases to free surface, a similar topographical change is also observed at the electrode edges when cycling a capacitor in ambient environment. The study suggests that surface electrochemical reactivity may have a significant impact on the functionality of this material in the ambient environment. However, even in the controlled atmosphere, the participation of the surface ions in polarization switching phenomena and ionic compensation is possible.

For over 50 years, ferroelectric materials have been actively explored in the context of potential information technology applications.[1-5] The existence of two or more polarization states that are switchable by an electric field combined with atomically thin widths of ferroelectric domain walls provides tremendous potential for high-density information storage. Correspondingly, multiple ferroelectric device geometries including ferroelectric gate field effect transistor (FeFET),[6] ferroelectric random access memories (FeRAM),[7] and ferroelectric tunneling barriers[8] have been explored theoretically and experimentally. However, the practical implementation of the ferroelectric-semiconductor memory has been limited by the thermodynamic incompatibility between the classical oxide perovskites and semiconductors such as silicon, germanium, and gallium arsenide.[9] Extensive efforts have been undertaken to achieve the kinetic stabilization of the ferroelectric-semiconductor interfaces via the introduction of interface layers.[10-12] However, these approaches are generally incompatible with semiconductor backend processing.

The paradigm shift in this field was achieved with the discovery of ferroelectricity in hafnia solid solutions.[13] In contrast to perovskites, these materials can be deposited using techniques such as atomic layer deposition and are directly compatible with semiconductor fabrication lines.[14] Accordingly, the last 5 years have seen exponential growth of research activity in this field. In addition to hafnia and zirconia, ferroelectric properties have been discovered in nitrides such as $Al_xSc_{1-x}N$ [15] and $Al_xB_{1-x}N$,[16] and oxides such $Zn_xMg_{1-x}O$.[17] However, ferroelectricity in the wurtzite compounds shows multiple unique behaviors that are uncommon in classical ferroelectrics. These include effects such as wake up, i.e., emergence of the ferroelectric response after multiple cycles,[17,18] the possible absence of a paraelectric phase,[19-21] retention of ferroelectricity down to small thicknesses (~ 10 nm)[22,23] and the ability to observe square ferroelectric hysteresis loops even in materials that are structurally disordered.

It is also important to mention that these behaviors are reminiscent of unusual behaviors observed in very thin films of classical ferroelectric or non-ferroelectric oxides such as $SiO_2$ or unstrained $SrTiO_3$ as summarized in Ref. [[24]]. In many cases, these materials systems exhibit a continuum of states with dissimilar electromechanical responses[25,26] and long-term relaxation[27,28] that cannot be ascribed solely to spontaneous polarization and domain state. For some materials, these behaviors can be attributed to the coupled ferroionic states emerging due to the interplay between the ferroelectricity and surface electrochemical processes.[29,30] Correspondingly, exploration of the new ferroelectric materials has to balance the novel bulk behavior and the potential reversible and irreversible electrochemical phenomena on open surfaces.

Here, strong irreversible electrochemical phenomena are observed on open surfaces of $Al_xB_{1-x}N$ thin films in which protrusion-shaped structure is formed by applying electric biases in ambient conditions. This electrochemical phenomenon is observed in both 20 nm and 205 nm thick films for both alternating current (AC) and direct current (DC) biases. It is found that negative voltage is the causal origin of this electrochemical phenomenon. It is proposed that the small protrusion-shaped structures produced are associated with field-assisted hydrolysis on the $Al_xB_{1-x}N$ surface.

Two $Al_xB_{1-x}N$ thin films synthesized on a sapphire substrate with W bottom electrodes and thicknesses of 20 nm ($Al_{0.93}B_{0.07}N$) and 205 nm ($Al_{0.93}B_{0.07}N$) were utilized for this work. These films show large reorientable polarizations exceeds 120 µC/cm$^2$, with coercive fields ranging between 5-6 MV/cm in metal-ferroelectric-metal structures, as shown in Figure 1a, agree with previous reports.[16] To establish the presence of ferroelectricity and switchable polarization on $Al_xB_{1-x}N$ open surfaces, we explore them via piezoresponse force microscopy (PFM) imaging and spectroscopy measurements. Typically, ferroelectric polarization can be switched by applying a DC bias greater than the coercive field to the scanning probe microscopy tip,[31-33] with subsequent PFM imaging to confirm polarization reversal. As such, we applied ± 10 V to the 20 nm thick $Al_{0.93}B_{0.07}N$ film to reverse the local polarization with subsequent PFM imaging. Interestingly, the application of bias during scanning lead to highly unstable imaging during bias applications, as shown in Figure S1. Similarly, post-bias scans have exhibited an obvious morphology change. It should be noted that there are multiple potential origins of scanning instabilities in PFM including tip degradation, sweeping of surface contaminates by the tip motion, deposition of the silane compounds from tip packing,[34] and electrochemical reactions[35] on the sample surface. However, the instabilities observed in this case were generally consistent with the intermittent surface reactivity, while the tip state remained stable and the contaminates transfer from tip to the surface or laterally along the surface remained minimal.

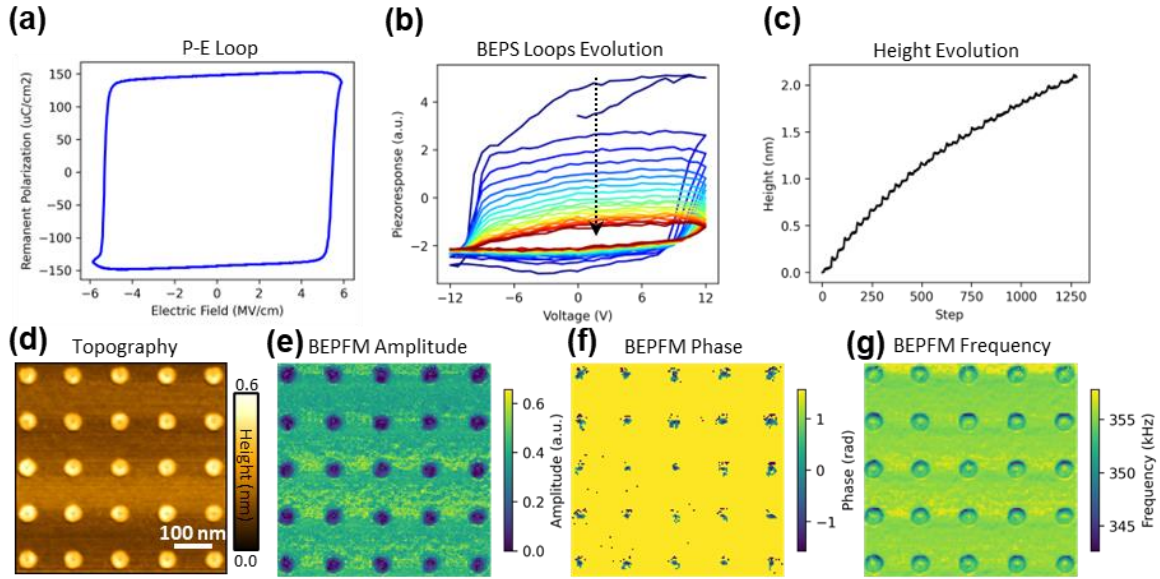

**Figure 2.** (a) Polarization-electric field hysteresis loop of a representative 200 nm thick $Al_{0.93}B_{0.07}N$ grown on W (110 orientation) on c plane $Al_2O_3$ substrate, the P-E measurement is performed at 1 kHz frequency with a triangular voltage waveform. (b) BEPS measurement of the 20 nm thick $Al_{0.93}B_{0.07}N$ film in ambient, averaged BEPS loops over 5×5 grid (500 x 500 nm$^2$ region), where the color of loops represents measurement step. The DC waveform for BEPS measurement is shown in Supplementary Information Figure S1. (c) the sample height evolution as a function of BEPS step. (d) topography after BEPS measurement. (e-g) BEPFM after BEPS measurement.

To further understand these behaviors, we have explored the evolution of the surface topography and polarization state using band excitation piezoresponse spectroscopy measurements (BEPS).[36] Specifically, a 5×5 spectroscopic array comprising 20 cycles of a triangular BEPS waveform with a magnitude of 12 V was applied to switch the polarization, as illustrated in Figure S2. The total duration of the triangular waveform is 12 seconds, note that this waveform is composed of both on-field and off-field conditions at each step. The averaged piezoresponse vs. $V_{dc}$ loops are shown in Figure 1b, where the blue and red colors represent the beginning and end of the waveform, respectively. At the beginning of the BEPS measurement, the loops show polarization switching, comparable to that of metal-ferroelectric-metal capacitors shown in Figure 1a. However, with subsequent cycles, the loops begin to shrink, and eventually remanent polarization drastically reduced with higher cycling. The sample morphology evolution during this process was tracked through the height channel, shown in Figure 1c, it indicates that the sample height increases during the BEPS cycling. The $Al_{0.93}B_{0.07}N$ thin film topography and band excitation (BE) PFM measurements after the BEPS measurement are shown in Figure 1c-f. Topography (Figure 1d) shows protruding structures in the areas where BEPS was performed, *i.e.,* where $V_{dc}$ was applied. Additionally, BEPFM imaging shows a lower piezoresponse (Figure 1e) and resonance frequency shift (Figure 1g) at the corresponding locations. These observations clearly indicate that application of the BEPS waveform on a $Al_{0.93}B_{0.07}N$ bare surface induces a surface phenomenon.

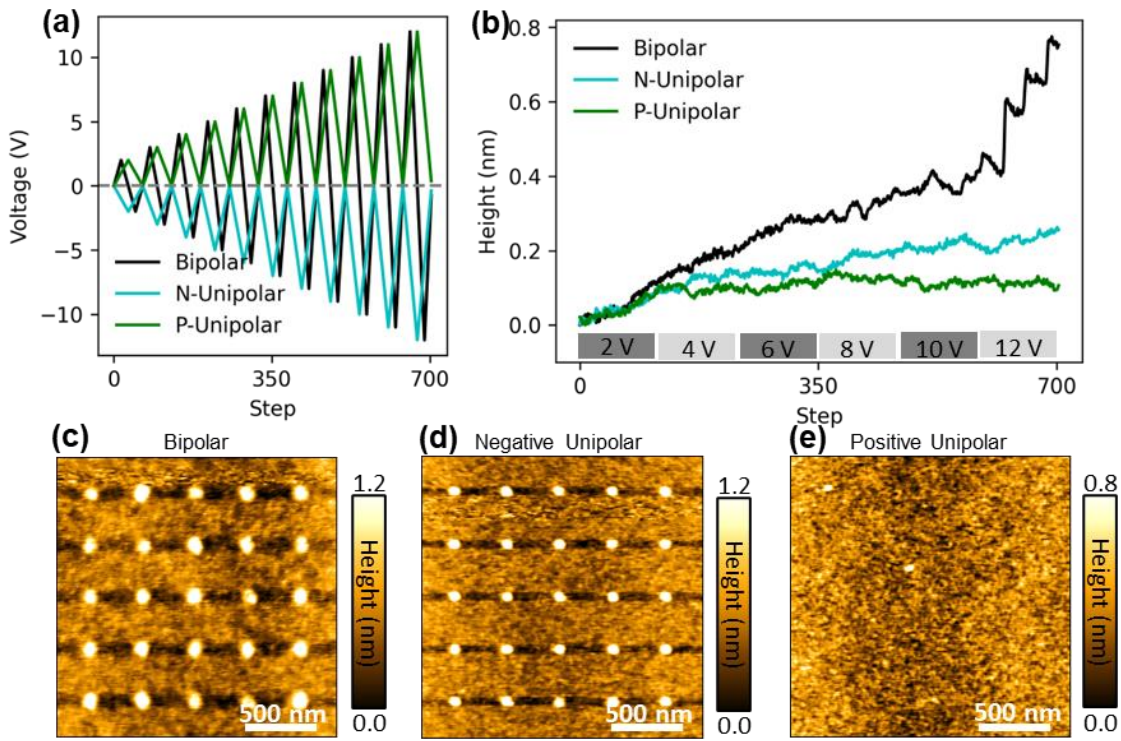

**Figure 2. Polarity effect**. (a) bi-polar, negative unipolar (N-unipolar), and positive unipolar (p-unipolar) FORC waveform for BEPS measurement. (c, d, e), topography change after respective FORC-BEPS measurement. BEPS response under respective FORC waveforms are shown in Figure S2.

To better understand the origins of the observed surface phenomena, bipolar and unipolar first order reversal curve measurements (FORC), were utilized, as shown in Figure 2. Previously this approach has been used to explore the electrochemical reactivity on the surfaces of the Li-ion conductors and oxygen conductive materials such as ceria;[37,38] in ferroelectrics, FORC curves also allow switching distributions to be assessed.[39,40] In bipolar FORC, a gradually increasing triangular waveform centered around 0 V is applied to the sample; however, in unipolar FORC, gradually increasing negative or positive triangular waveforms are applied to the sample (Figure 2a). Note that the pristine polarization can be flipped by applying positive bias. Figure 2b shows the sample height change during the application of bipolar and unipolar FORC waveform. The sample height increases most significantly under the application of bipolar FORC waveform, in contrast, the sample height is mostly constant under positive unipolar waveform. The piezoresponse vs. $V_{dc}$ loops for bipolar, negative unipolar, and positive unipolar waveforms are shown in Figure S3. The morphology of the area after FORC measurements was checked via AFM topography measurement, as shown in Figure 2c, 2d and 2e. Both bipolar and negative unipolar FORC induced the formation of protrusion-like features on the sample surface (Figure 2c and 2d), however, no change on morphology is found after applying positive FORC (Figure 2e). These observations suggest that the negative bias is most likely the culprit of electrochemical reaction on the sample surface.

Noteworthily, shown in Figure 2b, the sample height starts to increase at the beginning of FORC waveform, where the waveform magnitude is just 2 V, this voltage is much smaller than the coercive voltage. For a 20 nm thick $Al_{0.93}B_{0.07}N$ film, in principle a DC bias larger than ±10 V can reverse the polarization. Figure 2b also shows a faster height increase around 10-12 V. This raises a question whether the electrochemical reaction is associated with the ferroelectric polarization switching via DC and AC bias. Therefore, to deconvolute surface electrochemistry and polarization switching, a thicker 205 nm film with a coercive voltage of 129V was also investigated. Here, the effect of both DC and AC bias on the electrochemical phenomenon at the $Al_{0.93}B_{0.07}N$ surface was assessed. Shown in Figure 3a is the experiment sequence; the BEPS measurement was performed 3 times for the same area. In each BEPS measurement (over 10×10 grid), a triangular DC waveform was applied to the sample and a topography measurement was performed after each BEPS measurement to check the morphology change induced by DC bias (as a sign of electrochemical reaction). Then, after 3 BEPS measurements, a ramping up AC waveform (2-50 V, 350-430 kHz) was applied to further investigate the surface electrochemistry dependence on AC bias. Finally, a topography measurement was acquired to check the morphology change induced by AC waveform. Shown in Figure 3b is the BEPS DC waveform and BEPS piezoresponse vs. $V_{dc}$ loops.

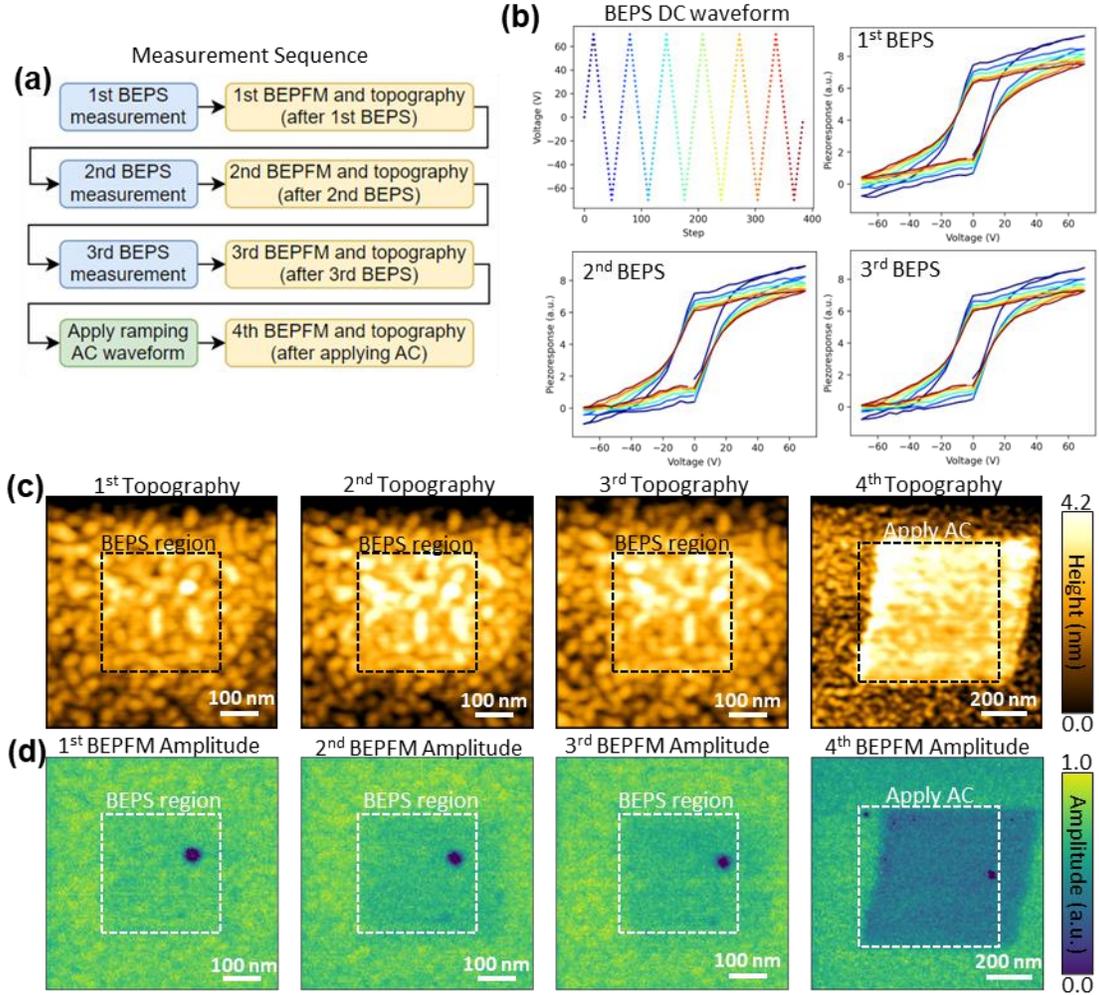

**Figure 3.** Electrochemical reaction without switching on a 205 nm $Al_xB_{1-x}N$ film. (a) Measurement sequence. (b) BEPS measurement results. (c) Topography changes induced by BEPS and BEAM measurements. (d) Piezoresponse changes induced by BEPS and BEAM measurement.

Shown in Figure 3c and 3d are the topography and BEPFM amplitude after each BEPS and BEAM measurement, respectively. It is clear that the topography change becomes more significant with increasing number of BEPS measurements, i.e., 1$^{st}$ – 3$^{rd}$ topography in Figure 3c. Along with the topography change, the piezoresponse amplitude also shows more significant decrease in the 1$^{st}$-3$^{rd}$ BEPFM amplitude maps in Figure 3d. The topography and piezoresponse changes after BEPS measurement are induced by DC bias, which are consistent with the results for the 20 nm Al$_{0.93}$B$_{0.07}$N film. Similar to the effect of DC waveform, AC waveform also induces topography and piezoresponse changes, as shown in the 4$^{th}$ topography and BEPFM amplitude in Figure 3c-d, suggesting the electrochemical reaction also occurs under AC waveform.

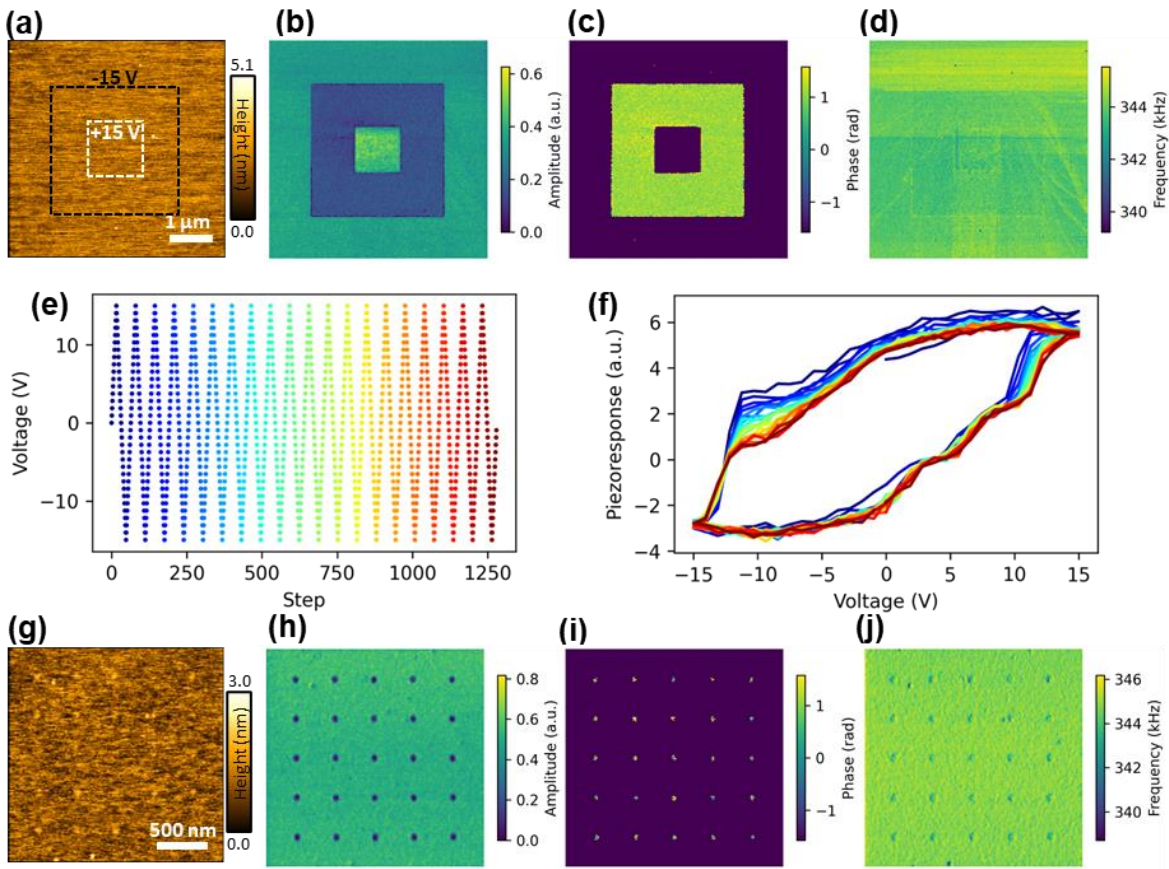

**Figure 4.** Experiment in Ar glovebox. (a-d) domain writing on Al$_x$B$_{1-x}$N thin film: (a) topography, (b-c) BEPFM amplitude, phase, and frequency, respectively. (e-f) BEPS measurement: (e) DC waveform for BEPS measurement and (f) averaged BEPS loops over 5×5 grids. (g-j) topography

and BEPFM after BEPS: (g) topography, (d-f) BEPFM amplitude, phase, and frequency, respectively.

For the bias dependent measurements in SPM, the surface chemistry can be sensitively affected by the humidity and the presences of adsorbed water layers.[41-44] To establish the role of humidity, similar domain writing and BEPS experiments were performed in an argon glovebox, where the $H_2O$ level is below 0.1 ppm, using the 20 nm thick film. Domain writing and BEPS results are shown in Figure 4. In Figure 4a-d, the local polarization state is reversed by applying ±15 V DC bias through the tip. Here, no obvious topography change is observed, in contrast to the domain writing in ambient atmosphere (Figure S1), where significant topography change is induced by DC bias. Figure 4e and 4f show the BEPS triangular DC waveform and loops results measured in the glovebox, the BEPS loops (20 cycles) are very stable in this case, in contrast to rapidly decreasing loop width when the measurement is performed in ambient (Figure 1b). These BEPS loops collected in a glovebox show some dispersion that can indicate the presence of electrochemical reactivity, the complex shape of the loop can also indicate a combination of normal switching and weakly-irreversible electrochemical processes. The topography and BEPFM response after the BEPS measurement are shown in Figure 4g-j, suggesting a finite electrochemical reaction when bias was applied to $Al_{0.93}B_{0.07}N$ in the glovebox. However, changes in the topography and piezoresponse are much weaker compared to those in ambient conditions. The electrochemical reaction occurred in glovebox is most likely due to the residue $H_2O$ absorbed on $Al_{0.93}B_{0.07}N$ surface.

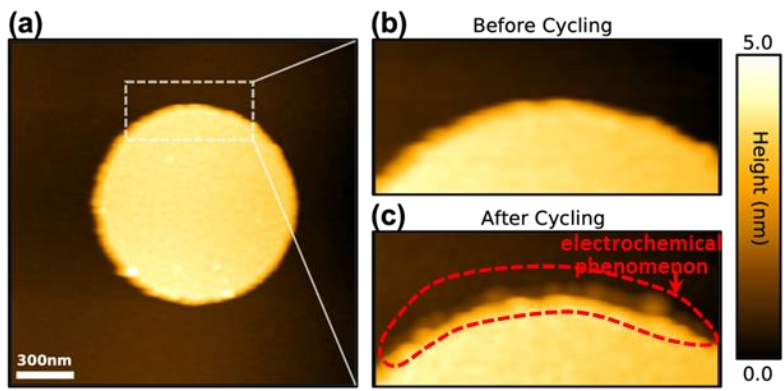

**Figure 5. Capacitor cycling.** (a) topography of 1μm capacitor on 20nm $Al_{0.93}B_{0.07}N$. (b) Magnified region located at top of capacitor illustrated in panel (b) before cycling, and (c) after cycling. Note, electrochemical reaction occurs at the boundary of capacitor after cycling (panel (c)).

To better understand chemical phenomena associated with observed effects, the protrusion-shaped structures were study using time-of-flight secondary ion mass spectrometry (ToF-SIMS). However, those studies didn't reveal any significant changes in the stoichiometry of switched regions, as shown in Figure S4. This may be due to low sensitivity of the technique and small size of the studied protrusion structures, which are smaller than size of the ion beam spot (~120 nm) used in the ToF-SIMS.

A hypothesis is that application of the negative bias to the tip in the presence of the water layer induces an electrochemical reaction with the increase of the molar volume of resulting product. Note that electrochemical reactions should always have reduction and oxidation half reactions at spatially separated sites, satisfying the conditions of detailed charge and mass balance. For electrochemical processes induced by the tip, the attribution of the cathodic and anodic processes can be ambiguous.[45] In analyzing the possible mechanisms, it is noted that aluminum and boron are unlikely to change their oxidation states, and the reaction product most likely is a mixture of aluminum and boron hydroxides. The alternative pathways can be either electrochemical oxidation of nitride to the $N_2$, or the splitting of the water molecules and simple field-assisted hydrolysis of the nitride surface. Based on the fact that the process is clearly water assisted and given the difficulty for electrochemical oxidation of nitride, likely mechanisms include:

Under tip: $Al_xB_{1-x}N + 3e^- + 6H_2O \Rightarrow (Al_xB_{1-x})(OH)_3 + NH_3 + 3OH^- + \frac{3}{2}H_2$

Away from tip: $2OH^- \Rightarrow H_2O + \frac{1}{2}O_2 + 2e^-$

Finally, we explore whether these phenomena are relevant for devices with top electrodes. Shown in Figure 5a are 1 μm diameter electrodes fabricated on the 20 nm thick $Al_{0.93}B_{0.07}N$. Here, we cycle the electrodes 100 times in both ambient and argon environments, similar to the bare surface measurements, while leveraging the SPM tip as an electrical contact to the capacitor. Before cycling, well defined capacitor edges are observed as shown in Figure 5b. However, after cycling in ambient, clear topographical changes can be observed at the capacitor edges as shown Figure 5c, indicating the electrochemical phenomenon is not isolated to only free surface measurements. Interestingly, these observations imply the functional response of $Al_{0.93}B_{0.07}N$ in small capacitor geometries is also susceptible to environmental effects, primarily isolated to the edges of the capacitor.

To summarize, polarization switching and domain dynamics in aluminum boron nitride have been explored using piezoresponse force microscopy and spectroscopies. It was found that in ambient environment the application of positive biases to the tip, either in the form of DC bias, or unipolar or bipolar FORC waveforms, or AC bias sweeps lead to significant surface expansion and decay of electromechanical response. This behavior is attributed to the tip-induced electrochemical reactions mediated by the presence of the water layer.

On transition to the controlled glovebox environment, the sharp polarization domain patterns can be created. However, hysteresis loops still exhibit dynamic changes with time and the change in surface topography can be observed on repeated cycling. Based on the shape of the hysteresis loop, we speculate that in this case the polarization switching, and the electrochemical surface changes coexist. Overall, we pose that electrochemical phenomena are in many cases inseparable part of polarization dynamics on open ferroelectric surfaces, due to the fact that polarization switching is possible only when the polarization charged is screened by electronic and ionic charges. In ambient and even controlled atmosphere environments when top electrodes are not present, the screening is ionic in nature and hence the surface undergoes multiple chemical changes during the application of periodic biases.

**Acknowledgements**

This effort (materials synthesis, PFM measurements) was supported as part of the center for 3D Ferroelectric Microelectronics (3DFeM), an Energy Frontier Research Center funded by the U.S. Department of Energy (DOE), Office of Science, Basic Energy Sciences under Award Number DE-SC0021118. The research (PFM measurements) was performed and partially supported at Oak Ridge National Laboratory's Center for Nanophase Materials Sciences (CNMS), a U.S. Department of Energy, Office of Science User Facility. ToF-SIMS characterization was conducted at the Center for Nanophase Materials Sciences, which is a DOE Office of Science User Facility, and using instrumentation within ORNL's Materials Characterization Core provided by UT-Battelle, LLC under Contract No. DE-AC05-00OR22725 with the U.S. Department of Energy.

**Conflict of Interest**

The authors declare no conflict of interest.